\documentclass[a4paper,11pt]{article}

\input{xy}
\xyoption{all}
\xyoption{poly}
\usepackage[all]{xy}
\usepackage{amsmath,amsthm,amsfonts,amssymb,stmaryrd,fancyhdr,mathtools}
\usepackage{latexsym}
\usepackage{color}
\usepackage{graphicx}
\usepackage{float}
\usepackage{fullpage}
\usepackage{mdframed}
\usepackage[font=small,labelfont=small]{caption}
\usepackage[T1]{fontenc}
\usepackage[utf8]{inputenc}
\usepackage{authblk}

\def\beq{\begin{equation}}
\def\eeq{\end{equation}}
\def\beqa{\begin{eqnarray}}
\def\eeqa{\end{eqnarray}}

\def\a{{\alpha}}
\def\b{{\beta}}
\def\g{{\gamma}}

\def\eps{{\epsilon}}

\def\m{{\mu}}
\def\n{{\nu}}

\newcommand{\cR}{{\cal R}}
\newcommand{\cS}{{\cal S}}
\newcommand{\cP}{{\cal P}}
\newcommand{\cQ}{{\cal Q}}
\newcommand{\cC}{{\cal C}}
\newcommand{\cM}{{\cal M}}
\newcommand{\cN}{{\cal N}}

\newcommand{\mR}{\mathbb{R}}

\title{Duality phases and halved maximal $D=4$ supergravity}
\author[1]{Sergio~M.~Iguri\thanks{siguri@iafe.uba.ar}}
\author[2]{Victor~A.~Penas\thanks{vpenas@iafe.uba.ar}}
\affil[1]{Instituto de Astronom\'{\i}a y F\'{\i}sica del Espacio (CONICET-UBA)

Casilla de Correo 67, Sucursal 28, 1428 Buenos Aires, Argentina}
\affil[2]{Departamento de F\'isica, Facultad de Ciencias Exactas y Naturales, 

Universidad de Buenos Aires,

Ciudad Universitaria, Pabell\'on I, 1428 Buenos Aires, Argentina}

\begin{document}

\maketitle

\begin{abstract}

The duality angles deformation developed by de Roo and Wagemans within the context of $N=4$ gauged supergravity is used in order to study certain classes of gaugings of $N=8$ supergravity, namely, those that are consistent when halving the maximal $D=4$ theory. After reviewing the truncation process from $N=8$ to $N=4$ supergravity in terms of the embedding tensor formalism, the de Roo-Wagemans phases method is implemented for solving the resulting constraints on the gauging parameters by means of the Sch\"on-Weidner ansatz. In contrast with the twenty semisimple $N=4$ gaugings admitting more than a single $SL(2)$ angle deforming their decompositions reported in the literature, it is proven that only three of them can be embedded back into the $N=8$ theory. The scalar potential derived for only two of these gauge groups exhibits an extremum in the origin of the scalar manifold. These extrema are not stable under fluctuations of all the scalar fields.

\end{abstract}

\bigskip



\newpage

\tableofcontents

\bigskip

\section{Introduction}

Recently, several authors \cite{Dibitetto:2011eu,Aldazabal:2011yz,Dibitetto:2012ia} have explored the possibility of consistently halving $D=4$ maximal gauged supergravities, namely, deformations that result from promoting a certain subgroup of the $N=8$ supergravity (electric/magnetic) duality group to a local invariance in a $4$-dimensional space-time. The main motivation for doing so is its relevance in type II superstring orbifold compactifications including $O$-planes and $D$-branes as well as both geometric and nongeometric background fluxes. Actually, while $O6/D6$ sources break half of the supersymmetries for type IIA superstring toroidal orientifold compactifications with gauge and geometric fluxes, leading thus to half-maximal supergravities in the low-energy regime, an embedding into a maximal supergravity theory could be consistent if the flux-induced tadpole for the Ramond-Ramond $7$-form that couples to the $O6/D6$ sources is canceled and if the twisted sector is projected out \cite{Dall'Agata:2009gv,Dibitetto:2011gm}. A similar situation can be observed when studying the untwisted sector of type IIB superstring toroidal orbifold compactifications with generic fluxes as in Ref.~\cite{Aldazabal:2008zza}. The interest was recently renewed within the context of double field theory since such reductions from $N=8$ to $N=4$ emerge as a consequence of some of the constraints of the formalism \cite{Aldazabal:2011nj,Geissbuhler:2011mx,Grana:2012rr,Dibitetto:2012rk}.

The so-called embedding tensor approach \cite{
deWit:2002vt,deWit:2005ub,deWit:2007mt,Schon:2006kz} is a formal scheme developed in order to describe all gauged supergravities in a unified way. Besides its success for classifying gauged supergravities based on strictly group-theoretical criteria, it has proven to be useful for the analysis of string theory realizations of both  $D=4$ maximal and half-maximal supergravity models \cite{Dall'Agata:2009gv,Aldazabal:2008zza,deWit:2003hq,Aldazabal:2010ef}. The contact between these two frameworks, the ``truncation'' from maximal to half-maximal supergravities, was advanced in Refs.~\cite{Aldazabal:2011yz,deWit:2003hq,Derendinger:2006jb} and it was fully developed within the embedding tensor formalism in Ref.~\cite{Dibitetto:2011eu}. Making explicit use of the branching rules of different representations of $E_{7(7)}$, the maximal duality group, under the action of the half-maximal global symmetry group $SL(2)\times SO(6,6)$, together with the linear and quadratic constraints on the embedding tensor of maximal supergravity \cite{deWit:2007mt}, it was found that, in addition to the already known constraints on the embedding tensor components of half-maximal supergravity, a new set of quadratic constraints must be imposed in order to guarantee a consistent reduction from $N=8$ to $N=4$ (see Fig.~1).

Leaving aside numerical computations \cite{Aldazabal:2011yz,Dibitetto:2011gm}, the only known semianalytical method to systematically solve the gauging constraints in the $N=4$ theory is derived from the duality phases deformation developed by de Roo and Wagemans in Ref.~\cite{deRoo:1985jh} for semisimple local symmetry groups. The formalism was further studied in Refs.~\cite{deRoo:2002jf,deRoo:2003rm}. According to this method, different angles are introduced for each simple factor of the gauge group in order to parametrize the couplings between the vector multiplets and the $SL(2)$ scalar fields of the theory. The emergence of the de Roo-Wagemans phases in the embedding tensor formalism has been investigated in Refs.~\cite{Schon:2006kz,Roest:2009dq}. Within this framework, the duality angles appear when, instead of realizing the gauging parameters strictly as structure constants, an ``unpolarized'' decomposition is considered. This unpolarization becomes crucial for moduli stability. If the $SL(2)$ angles are all equal, the corresponding gauging reduces to a purely electrical one, and the resulting scalar potential necessarily presents runaway directions.

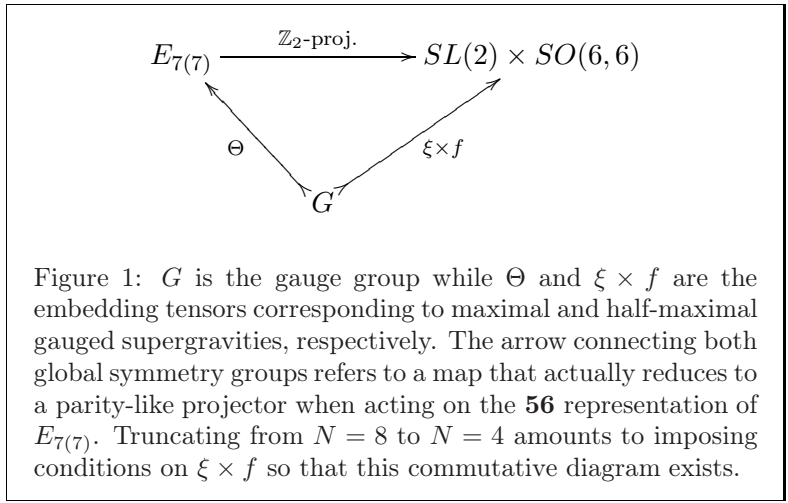
\vspace{0.3cm}

\begin{figure}[h]
\label{figura}
\centering
\begin{minipage}[b]{0.65\linewidth}
\begin{mdframed}
\begin{displaymath}
\xymatrix@R=40pt{
E_{7(7)}
\ar@{->}[rr]^-{\mathbb{Z}_2\textrm{-proj.}}
\ar@{<-<}[dr]_-{\Theta}
&& SL(2)\times SO(6,6)
\ar@{<-<}[dl]^-{\xi\times f} \\
& ~~G~ } \\ 
\end{displaymath}
\caption{$G$ is the gauge group while $\Theta$ and $\xi\times f$ are the embedding tensors corresponding to maximal and half-maximal gauged supergravities, respectively. The arrow connecting both global symmetry groups refers to a map that actually reduces to a parity-like projector when acting on the {\bf 56} representation of $E_{7(7)}$. Truncating from $N=8$ to $N=4$ amounts to imposing conditions on $\xi\times f$ so that this commutative diagram exists.}
\end{mdframed}
\end{minipage}
\end{figure}

The aim of this paper is to investigate those gaugings of maximal gauged supergravity that remain as such after a truncation halving the number of supersymmetries is performed, using the general construction of de Roo and Wagemans for solving the full set of constraints on the deformation parameters. We shall prove that no semisimple local groups except, eventually, those decomposable in strictly four $3$-dimensional simple groups, give rise to an unpolarized gauging. This fact strongly restricts the list of groups admitting a nontrivial stability analysis from the twenty reported in Ref.~\cite{deRoo:2003rm} to five. From these last ones, two must be also discarded due to the specific duality angles imposed by the truncation constraints. Namely, within the framework of the de Roo-Wagemans formalism, the impact of a consistent embedding back into the maximal theory reduces the number of gauge groups that could have at least two different $SL(2)$ phases to $15\%$. An interesting feature of the previous analysis is the fact that for gaugings arising from the duality angles deformation, one of the truncation constraints becomes redundant. A further study of the scalar potential shows that it does not have any extremum for one of the allowed groups, leaving us with just two gaugings exhibiting extrema in their potentials with partial stability under fluctuations of the scalar fields.

Let us point out that we have focused our attention on semisimple gaugings, which are those modelling the duality phases deformation, in order to systematize our treatment along the lines of Refs.~\cite{deRoo:2002jf,deRoo:2003rm}. A methodical way for studying nonsemisimple gaugings in the de Roo-Wagemans formalism is lacking, although some examples can be found in the literature \cite{Schon:2006kz}. We thus stress that our analysis is most certainly not exhaustive. In fact, many relevant gaugings coming, for instance, from a Scherk-Schwarz generalized dimensional reduction \cite{Andrianopoli:2002mf} or from type II superstring orientifold compactifications with fluxes \cite{Frey:2002hf,Angelantonj:2003up} do not necesarilly correspond to semisimple groups.

The paper is organized as follows. In section 2 we briefly review the basics of the embedding tensor formalism and the constraints on the gauging parameters in order to truncate from maximal to half-maximal $D=4$ supergravity. In section 3 we implement the de Roo-Wagemans method for solving these constraints. The impossibility of a nonpurely electric semisimple gauging when less than four simple factors decompose the gauge group is proven, and the list of allowed groups is systematically studied in order to get a solution involving different duality phases. The existence of extrema for these solutions and their stability under fluctuations of all scalar fields is addressed. In a final section, we collect our conclusions.

\section{From maximal to half-maximal supergravity}


Maximal supergravity can be only deformed by promoting some subgroup $G$ of the duality group $E_{7(7)}$ to a local symmetry, namely, by applying gaugings. The parametrization of all possible gaugings can be encoded in a single spurionic object transforming under the global symmetry group, the resulting embedding tensor being thus group-theoretically characterized. This embedding tensor belongs to the ${\bf 56} \times {\bf 133}$ representation of $E_{7(7)}$ and it determines the way in which the generators $X_{\cM}$ of the gauge group decompose in terms of the $E_{7(7)}$ generators $t_I$. Explicitly, the embedding parameters define a real tensor $\Theta_{\cM}{}^{I}$, $\cM=1,\dots,56$ and $I=1,\dots,133$ indexing the fundamental and the adjoint representations of $E_{7(7)}$, respectively, such that
\beq
X_{\cM}=\Theta_{\cM}{}^{I}t_I.
\eeq
The tensor $\Theta_{\cM}{}^{I}$ acts as a projector whose rank equals the dimension of the gauge group which must be less than or equal to $28$.

An admissible embedding tensor must satisfy a set of linear and quadratic constraints in order to ensure that the corresponding supergravity action remains supersymmetric after gauging and that the gauge group is actually a proper subgroup of $E_{7(7)}$. In order to truncate from maximal to half-maximal supergravity it is a suitable choice to deal with the restrictions these constraints imply on the charges associated with the particular gauging in hand instead of treating them directly on the embedding tensor components. These charges are defined by $X_{\cM\cN}{}^\cP=\Theta_{\cM}{}^I[t_I]_{\cN}{}^\cP$  and they act as structure constants of the local symmetry group. In fact, the commutation relations of the gauge group read
\beq
\label{gauge_algebra}
[X_{\cM},X_{\cN}]=-X_{\cM\cN}{}^{\cP} X_{\cP}.
\eeq

The linear constraints amount to restricting the embedding tensor to the ${\bf 912}$ representation of $E_{7(7)}$. Once the projection from the ${\bf 56} \times ({\bf 56} \times {\bf 56})_s$ representation of $E_{7(7)}$ on ${\bf 56} \times {\bf 133}$ is performed, the one on ${\bf 912}$ implies the following constraints for the charges \cite{deWit:2007mt}:
\begin{eqnarray}
\label{linconst}
X_{\cM[\cN\cP]}&=&0, \\
\label{linconst2}
X_{(\cM\cN\cP)} &=&0, \\
\label{linconst3}
X_{\cM\cN}{}^{\cM}&=&0.
\end{eqnarray}
In Eqs.~(\ref{linconst})-(\ref{linconst3}) we have denoted $X_{\cM\cN\cP} = X_{\cM\cN}{}^{\cQ} \Omega_{\cQ\cP}$, where $\Omega_{\cQ\cP}$ is the $\textrm{Sp}(56,\mR)$ invariant skew-symmetric matrix, used to raise and lower fundamental indices. After these linear constraints are imposed, the quadratic ones reduce to
\beq
\label{QC8} 
X_{\cM \cN \cP} X_{\cQ \cR \cS} \Omega^{\cM \cQ} = 0.
\eeq 

The strategy for carrying on the truncation from maximal to half-maximal supergravity requires to specify the charges $X_{\cM \cN \cP}$ strictly in terms of the embedding tensor parameters corresponding to $N=4$ gauged supergravity and then realize the equations (\ref{linconst})-(\ref{QC8}) as constraints of the half-maximal theory. In order to do so, the branching rules of different $E_{7(7)}$ representations under the action of $SL(2)\times SO(6,6)$ should be used before a projection halving the number of supersymmetries is performed. The decomposition ${\bf 56} \rightarrow ({\bf 2},{\bf 12}) + ({\bf 1},{\bf 32})$ is of particular relevance. It amounts to the index splitting $\cM = (\a,M) \oplus \m$, where $\alpha=\pm$ is a $SL(2)$ index, $M=1,\dots,12$ is a $SO(6,6)$ vector index and $\m=1,\dots, 32$ is a Majorana-Weyl left-handed fermionic index of $SO(6,6)$. The discrete $\mathbb{Z}_2$-projection, that corresponds to orientifolding the model in string theory realizations of maximal supergravity, results in a parity acquirement for every index: while the bosonic indices $\a$ and $M$ become even, the fermionic indices become odd. Only states which are even truncate from maximal to half-maximal supergravity \cite{Derendinger:2006jb} so that the skew-symmetric matrix $\Omega_{\cM\cN}$ becomes block-diagonal. Its non trivial components are:
\begin{eqnarray}
\label{Omega}
\Omega_{\a M \b N} &= &\eps_{\a \b} \eta_{MN}, \\
\label{Omega2}
\Omega_{\m \n} &= &\mathcal{C}_{\m \n},
\end{eqnarray}
where $\epsilon_{\a\b}$ is the $2$-dimensional Levi-Civita symbol associated to the $SL(2)$ factor, $\eta_{MN}$ is the $SO(6,6)$ metric and $C_{\m\n}$ is the charge conjugation matrix of $SO(6,6)$.

The representation ${\bf 912}$ of $E_{7(7)}$ decomposes as ${\bf 912} \rightarrow ({\bf 2},{\bf 12}) + ({\bf 2},{\bf 220}) + ({\bf 1},{\bf 352'}) + ({\bf 3},{\bf 32})$. The components of the embedding tensor sitting in $({\bf 1},{\bf 352'})$ and $({\bf 3},{\bf 32})$ are odd, so that they are projected to zero, allowing to write the charges $X_{\cM \cN \cP}$ in terms of the half-maximal supergravity embedding tensor parameters $\xi_{\a M} \in ({\bf 2},{\bf 12})$ and $f_{\a MNP} = f_{\a [MNP]} \in ({\bf 2},{\bf 220})$, which are purely bosonic.

Once the most general ansatz for the charges compatible with the symmetry in their last two indices is considered and the linear constraints are imposed, one gets:
\begin{eqnarray}
\label{X1}
X_{\a M \b N \g P} &=& - \epsilon_{\b \g} f_{\a MNP} - \epsilon_{\b \g} \eta_{M [N} \xi_{ \a P]} -\epsilon_{\a (\b} \xi_{\g) M} \eta_{NP}, \\
\label{X2}
X_{\a M \m \n} &=& -\frac{1}{4} f_{\a MNP} \left[ \g^{NP} \right]_{\m \n}  -  \frac{1}{4} \xi_{\a N} \left[ \g_{M}{}^{N} \right]_{\m \n}, \\
\label{X3}
X_{\m \a M \n} &=& X_{\m \n \a M} = - \frac{1}{8} \xi_{\a M} \cC_{\m \n} + \frac{1}{8} \xi_{\a N} \left[ \g_{M}{}^{N} \right]_{\m \n}  \nonumber \\
&+& \frac{1}{8} f_{\a MNP} \left[ \g^{NP} \right]_{\m \n} - \frac{1}{24} f_{\a NPQ} \left[ \g_{M}{}^{NPQ} \right]_{\m \n},
\end{eqnarray}
where  $[\gamma^M]^{\m\dot{\nu}}$ and $[\bar{\gamma}^M]_{\mu\dot{\n}}$ are the $32\times 32$ matrix blocks that appear in the decomposition of the Dirac matrices of $SO(6,6)$. Gamma matrices with more than one index refer to antisymmetrized products of $[\gamma^M]$.

By plugging Eqs.~(\ref{Omega})-(\ref{X3}) into Eq.~(\ref{QC8}), the maximal supergravity quadratic constraints are expressed in terms of the half-maximal supergravity embedding parameters. One obtains:
\begin{eqnarray}
\label{QC41} 
\xi_{\a M} \xi_{\b}{}^{M} &=& 0, \\
\label{QC42}
\xi_{(\a}{}^{P} f_{\b)MNP} &=& 0, \\
\label{QC43}
3f_{\a R[MN} f_{\b PQ]}{}^{R}+2\xi_{(\a [M}f_{\b)NPQ]} &=& 0, \\
\label{QC44}
\eps^{\a \b}\left(\xi_{\a}{}^{P}f_{\b MNP}+\xi_{\a M}\xi_{\b N}\right) &=& 0, \\
\label{QC45}
\eps^{\a \b}\left[f_{\a RMN}f_{\b PQ}{}^{R}-\xi_{\a}{}^{R}f_{\b R[M[P}\,\eta_{Q]N]}-\xi_{\a [M}f_{\b N]PQ}+ f_{\a MN[P}\xi_{\b Q]}\right] &=& 0, \\
\label{QC46}
f_{\a MNP} f_{\b}{}^{MNP} &=& 0, \\
\label{QC47}
\left. \eps^{\a \b} f_{\a [MNP} f_{\b QRS]}  \right|_{\textrm{SD}} &=& 0,
\end{eqnarray}
where the subindex ``$\textrm{SD}$'' in the last equation stands for the self-dual part of the $SO(6,6)$ $6$-form.

Constraints (\ref{QC41})-(\ref{QC45}) can be recognized as the quadratic constraints of half-maximal supergravity found in Ref.~\cite{Schon:2006kz}. The additional constraints (\ref{QC46}) and (\ref{QC47}) define the subset of $N=4$ gaugings that are consistent with an embedding back into the $N=8$ theory \cite{Dibitetto:2011eu}. 

\section{Duality phases in halved maximal supergravity}

\subsection{Purely electric gaugings}

In the particular frame in which both electric and magnetic fields transform as vectors under the action of $SO(6,6)$, purely electric gaugings are reached by setting $f_{-MNP}=0$ and $\xi_{\a M}=0$, so that only $f_{+MNP}$ is non vanishing. In this case, equations \eqref{QC41}-\eqref{QC47} read
\begin{eqnarray}
\label{QC43bis}
f_{+ R[MN} {f_{+ PQ]}}^{R} &=& 0, \\
\label{QC43bis2}
f_{+ MNP} f_{+}{}^{MNP} &=& 0,
\end{eqnarray}
and, in addition, there is a linear constraint on $f_{+MNP}$, namely, $f_{+MNP}=f_{+[MNP]}$.

Let us first concentrate on Eq.~(\ref{QC43bis}). Since it reduces to a Jacobi-like identity, the most natural attempt for solving it is to identify some of the embedding parameters directly as structure constants of a gauge group $G$. Moreover, a semisimple choice automatically ensures the validity of the linear constraint. Nevertheless, in order to properly realize this group as a local symmetry, its Lie algebra must be embedded into the space of electric vector fields in such a way that the preimage of $\eta_{MN}$ equals, up to a global factor, the Killing-Cartan form of $G$. While the absolute value of this factor can be absorbed by redefining the generators of the gauge group, this condition puts a restriction on the signature of its metric. Once this metric compatibility is fulfilled, an explicit solution for Eq.~\eqref{QC43bis} is obtained by defining $f_{+MNP}$ as an extension to zero of the $3$-form associated with the structure constants of $G$. The possible simple groups that can appear as factors in $G$ are $SO(3)$, $SO(2,1)$, $SO(3,1)$, $SL(3, \mathbb{R})$, $SU(2,1)$, $SO(4,1)$ and $SO(3,2)$.

Another related but, in principle, different solution of Eq.~(\ref{QC43bis}) that will be useful in the next subsections can be obtained when the dimension of $G$ is less than or equal to $6$ if, instead, the structure constants $3$-form is effectively Hodge-dualized relative to a predefined $6$-dimensional subspace and then extended to the whole fundamental representation of $SO(6,6)$. Explicitly, let $f_{MNP}$ be the trivial extension of the $3$-form associated with the structure constants of $G$ to a given $6$-dimensional subspace of electric vector fields and consider its Hodge-dual, namely, $*f_{MNP}=(3!)^{-1}\epsilon_{MNPQRS}f^{QRS}$, where $\epsilon_{MNPQRS}$ is the $6$-dimensional Levi-Civita symbol. Metric compatibility is again assumed. We have
\beq
*f_{RMN}{*f}_{PQ}{}^{R} = \pm \frac{10}{3} \eta_{P[M} f_{RST} \eta_{N]Q} f^{RST},
\eeq
where the global sign depends on the signature of the $6$-dimensional metric. This expression can be rewritten as
\beq
*f_{RMN}{*f}_{PQ}{}^{R} = \pm \left[ \frac{1}{3} f_{RST} \eta_{M[P} \eta_{Q]N}f^{RST} + 2 f_{RS[M} \eta_{N][P} f^{RS}{}_{Q]} + f_{RMN} f_{PQ}{}^{R} \right],
\eeq
from where it follows that
\beq
*f_{R[MN}{*f}_{PQ]}{}^{R} = \pm f_{R[MN} f_{PQ]}{}^{R} = 0,
\eeq
namely, the dual $3$-form $*f_{MNP}$ satisfies the Jacobi identity as long as $f_{MNP}$ does. A solution for Eq.~\eqref{QC43bis} is consequently obtained by defining $f_{+MNP}$ as a trivial extension of $*f_{MNP}$. Let us point out that although this procedure could be insubstantial in the semisimple case, in the sense that it can bring gaugings simply related by a change of basis, it provides well defined gaugings even if the local symmetry group is not semisimple. Indeed, while in this case $f_{MNP}$ does not necessarily equal $f_{[MNP]}$, the form $*f_{MNP}$, defined now as the Hodge-dual of the totally antisymmetric part of $f_{MNP}$, satisfies the linear constraint.

Notice that none of the $3$-forms introduced above is a solution for Eq.~(\ref{QC43bis2}) when the gauge group is strictly simple since, in this case, $f_{+ MNP} f_{+}{}^{MNP}$ is proportional to the dimension of $G$. As before, the absolute value of the proportionality constant is irrelevant. On the contrary, its sign depends on the embedding of $G$ into the fundamental representation of $SO(6,6)$ and eventually it could bring a way to generate solutions for both Eqs.~(\ref{QC43bis}) and (\ref{QC43bis2}) in a more general case. Let us illustrate this fact with a particular example. Let us assume that $G$ is a semisimple group that can be decomposed as $G^{(1)}\times G^{(2)}$, $G^{(1)}$ and $G^{(2)}$ having the same dimension, and consider the structure constants $3$-forms of both factors. It is straightforward to see that the (direct) sum of the trivial extensions of these $3$-forms provides a solution to (\ref{QC43bis}). In addition, it could correspond to a solution of Eq.~(\ref{QC43bis2}) as well if the preimage of $\eta_{MN}$ under the embedding of $G^{(1)}$ differs in a sign from that under the embedding of $G^{(2)}$, since in this case the contributions to $f_{+ MNP} f_{+}{}^{MNP}$ coming from both $3$-forms cancel.

Even when we are able to find solutions to constraints (\ref{QC43bis}) and (\ref{QC43bis2}), it is known that purely electric gaugings do not stabilize all moduli and therefore de Roo and Wagemans introduced a deformation of the theory in Ref.~\cite{deRoo:1985jh}, starting from a semisimple gauge group as before but introducing further a phase for every of its simple factors as an additional parameter in the description of the corresponding gauging. 

Before discussing the de Roo-Wagemans method, let us recall that, besides all semisimple choices, there could be many other nonsemisimple solutions to the constraints. We mention as an example the $\textrm{U}(1)^3$ gauging referred to in Ref.~\cite{Schon:2006kz}. Setting $u$, $v$ and $w$ to be three mutually orthogonal linearly independent lightlike vectors, the components of the volume form $u_{[M}v_Nw_{P]}$ define a solution of Eqs.~(\ref{QC43bis}) and (\ref{QC43bis2}). Further generalizations of this case to $3$-forms with a lightlike domain, corresponding all to Abelian gaugings, prove to be also solutions of both constraints \eqref{QC43bis} and (\ref{QC43bis2}). 

\subsection{The de Roo-Wagemans phases}

The de Roo-Wagemans deformation constitutes the only semianalytical approach when looking for solutions of constraints \eqref{QC41}-\eqref{QC45}. Here we will make use of the Sch\"on-Weidner ansatz \cite{Schon:2006kz}, which implements the duality phases method within the embedding tensor formalism, in order to explore solutions to the extended system \eqref{QC41}-\eqref{QC47} when $\xi_{\a M}=0$. Under this assumption, the system reduces to
\begin{eqnarray}
\label{QC43bb}
f_{\a R[MN}f_{\b PQ]}{}^{R} &=& 0, \\
\label{QC43bb2}
\eps^{\a \b} f_{\a RMN}f_{\b PQ}{}^{R} &=& 0, \\
\label{QC46bb}
f_{\a MNP} f_{\b}{}^{MNP} &=& 0, \\
\label{QC46bb2}
\left. \eps^{\a \b} f_{\a [MNP} f_{\b QRS]}  \right|_{\textrm{SD}} &=& 0.
\end{eqnarray}

Let us consider a decomposition of the Lie algebra of the gauge group $G$ into $K$ mutual orthogonal subspaces so that, for a general vector $v_M$ we have
\begin{equation}
v_M = \sum_{i=1}^K \pi^{(i)}_{M}{}^N v_N,
\end{equation}
with
\begin{equation}
\eta^{MP} \pi^{(i)}_{M}{}^N \pi^{(j)}_{P}{}^Q = 0 \qquad \text{for} \qquad i \neq j,
\end{equation}
where $\pi^{(i)}_{M}{}^N$, $i=1 \ldots K$, correspond to the orthogonal projectors onto each subspace.

As before, let us consider the structure constants of $G$ defining, by trivial extension, a $3$-form $f_{MNP}$, antisymmetric in its three indices, i.e., $f_{MNP}=f_{[MNP]}$, and satisfying the identity $f_{R[MN}f_{PQ]}{}^{R}=0$. Moreover, let us assume the decomposition of $G$ to be such that $f_{MNP}$ does not mix between the subspaces; namely, $f_{MNP}$ decomposes into a sum of independent $3$-forms nontrivially defined on each subspace:
\begin{equation}
\label{refere}
f_{MNP} = \sum_{i=1}^K f^{(i)}_{MNP},
\end{equation}
where
\begin{equation}
\label{refere2}
f^{(i)}_{MNP} = \pi^{(i)}_{M}{}^Q \pi^{(i)}_{N}{}^R \pi^{(i)}_{P}{}^S f_{QRS}.
\end{equation}

This implies, in turn, that the gauge group splits into $K$ factors $G=G^{(1)} \times G^{(2)} \times \ldots \times G^{(K)}$,
$f^{(i)}_{MNP}$ being the extension of the structure constants $3$-form associated with the $i$th factor, each of them satisfying separately the Jacobi identity. Even when in the semisimple case this decomposition of $G$ is naturally associated with its decomposition into simple factors, we stress that the above construction could apply for non semisimple gaugings as well. 

Solutions of the constraints (\ref{QC43bb}) and (\ref{QC43bb2}) in terms of $f^{(i)}_{MNP}$ are found to be generally given by the Sch\"on-Weidner ansatz \cite{Schon:2006kz}:
\begin{equation}
f_{\alpha MNP} = \sum_{i=1}^K w^{(i)}_\alpha f^{(i)}_{MNP},
\label{GaugingRW}
\end{equation}
where the $w^{(i)}$ are arbitrary ${\rm SL}(2)$ vectors that we could restrict to have unit length without loss of generality, i.e.,
\begin{equation}
w^{(i)}  = ( w^{(i)}_+ , \, w^{(i)}_- ) = ( \cos \alpha_i ,  \, \sin \alpha_i ).
\end{equation}
The $\alpha_i \in \mathbb{R}$, $i=1\ldots K$, are the so-called duality angles first introduced by de Roo and Wagemnans in Ref.~\cite{deRoo:1985jh}. Solution (\ref{GaugingRW}) is given up to ${\rm SL}(2)$ transformations so that we can freely assume that the first duality angle $\alpha_1$ vanishes, namely, we can set $w^{(1)} = (1,0)$, proving that cases with $K=1$ are always equivalent to purely electric gaugings.

It is convenient to emphasize that, in order to realize the gauge group in view of Sch\"on-Weidner ansatz, it is mandatory, as it was before in the purely electrical case, to embed the Lie algebra of each $G^{(i)}$ into the electric vector fields space in such a way that the preimage of $\eta_{MN}$ agrees, up to a real factor, with the associated Cartan-Killing form $\eta^{(i)}_{MN}$. Again, the absolute value of this factor is irrelevant but its sign could restrict the subgroups of $G$ allowed to contribute in Eq.~(\ref{refere}) in order to preserve the global signature of the metric $\eta_{MN}$.

In the next subsection we shall analyze under which conditions the Sch\"on-Weidner ansatz provides a solution for the remaining quadratic constraints (\ref{QC46bb}) and (\ref{QC46bb2}).

\subsection{Solving the truncation constraints}

Let us focus on semisimple gaugings, for which there is a natural decomposition of the space spanned by $\left\{v^M\right\}$ into mutually orthogonal subspaces so that $K$ can be freely assumed to equal the number of simple factors in $G$.

When replaced in Eqs.~(\ref{QC46bb}) and (\ref{QC46bb2}), the Sch\"on-Weidner ansatz (\ref{GaugingRW}) gives
\begin{align}
\label{QC46bbb}
&\sum_{i,j=1}^K w^{(i)}_\alpha w^{(j)}_\beta f^{(i)}_{MNP} f^{(j)MNP} = 0, \\
\label{QC47bbb}
& \sum_{i,j=1}^K \alpha_{ij} \left. f^{(i)}_{[MNP} f^{(j)}_{QRS]}  \right|_{\textrm{SD}} = 0,
\end{align}
where $\a_{ij}=\eps^{\a \b}w^{(i)}_\alpha w^{(j)}_\beta = \sin{(\a_j-\a_i)}$. Notice that $\a_{ii}=0$. The double sum in Eq.~(\ref{QC46bbb}) is reduced to a single one by virtue of the orthogonality of the nontrivial domains of the $3$-forms involved. On the other hand, we can halve the number of summands in Eq.~(\ref{QC47bbb}) using the symmetry of every term under the permutation of the block indices. We explicitly get
\begin{align}
\label{QC46bbbo}
&\sum_{i=1}^K w^{(i)}_\alpha w^{(i)}_\beta \left|f^{(i)}\right|^2 = 0, \\
\label{quad3}
&\sum_{i<j=1}^K \a_{ij} f^{(i)}_{[MNP} f^{(j)}_{QRS]} + \frac{1}{6!} \, \sum_{i<j=1}^K \a_{ij} \epsilon_{MNPQRSTUVWXY} f^{(i)TUV} f^{(j)WXY} = 0,
\end{align}
where $\epsilon_{MNPQRSTUVWXY}$ is the $12$-dimensional Levi-Civita symbol, associated with $SO(6,6)$, and we have introduced the notation  $\left|f^{(i)}\right|^2=f^{(i)}_{MNP} f^{(i)MNP}$.

Let us analyze the highly overdetermined system contained in Eq.~(\ref{quad3}). In order to solve it, we shall consider the subsystem obtained by contracting it with $f^{(k)MNP} f^{(l)QRS}$ for every pair of block indices $k<l$. Using the orthogonality of the spaces where the $3$-forms effectively act we get
\beq
f^{(k)MNP} f^{(l)QRS}f^{(i)}_{[MNP} f^{(j)}_{QRS]}= \frac{3!^2}{6!} \, \delta_{ik}\delta_{jl} \left|f^{(k)}\right|^2 \left|f^{(l)}\right|^2,
\eeq
so that Eq.~(\ref{quad3}) is rewritten as
\begin{eqnarray}
\label{quad4}
\a_{kl} \left|f^{(k)}\right|^2 \left|f^{(l)}\right|^2 +  \frac{1}{3!^2} \sum_{i<j=1}^K \a_{ij} \epsilon_{MNPQRSTUVWXY} f^{(i)MNP} f^{(j)QRS} f^{(k)TUV} f^{(l)WXY} = 0.
\end{eqnarray}

Since $\left|f^{(k)}\right|^2$ is proportional to the dimension of $G^{(k)}$, the coefficients appearing in the first term of the left-hand side of Eq.~(\ref{quad4}) are always different from zero. On the other hand, the contraction of the structure constants $3$-forms with the Levi-Civita symbol in the second term is antisymmetric under the interchange of any pair of block indices, so that it vanishes if any index is repeated. This fact implies that there is no contribution to Eq.~(\ref{quad4}) coming from this term if the decomposition of $G$ involves less than four factors. In this case, it follows that $\a_{kl}=0$ for every pair $k<l$, {\em i.e.}, $\alpha_k=\a_l+n\pi$ with $n \in \mathbb Z$, which shows that $f_{+MNP}$ and $f_{-MNP}$ are proportional. After performing a $SL(2)$ transformation in such a way that $\a_1$ is set to zero one concludes that semisimple gaugings are consistent with a truncation from maximal to half-maximal supergravity only in the purely electrical case except, eventually, for the specific situation in which the gauge group is decomposed in four or more simple factors.

After checking the list of simple subgroups of $SO(6,6)$ it should be clear that non purely electric gaugings can only be achieved if strictly four $3$-dimensional simple factors, namely, $SO(3)$ and/or $SO(2,1)$, decompose $G$. We list all allowed semisimple groups in Table 1. The signs in the last four columns indicate how the embedding of each group into the fundamental representation of $SO(6,6)$ is realized with respect to the relative signs between the corresponding Killing-Cartan form and the metric $\eta_{MN}$. The sign $+$ is used when positive entries are associated with the compact directions and negative entries with the noncompact ones, the sign $-$ is used otherwise.

From Eq.~(\ref{quad4}) and for all groups in Table \ref{tabla1}, we obtain a homogeneous system with six equations coupled in pairs, namely,
\begin{eqnarray}
\label{sys1}
 \a_{ij} \left|f^{(i)}\right|^2 \left|f^{(j)}\right|^2 + \frac{1}{3!^2} \, \a_{kl} \epsilon_{MNPQRSTUVWXY} f^{(i)MNP} f^{(j)QRS} f^{(k)TUV} f^{(l)WXY} = 0,
\end{eqnarray}
\begin{eqnarray}
\label{sys2}
 \frac{1}{3!^2} \, \a_{ij} \epsilon_{MNPQRSTUVWXY} f^{(i)MNP} f^{(j)QRS} f^{(k)TUV} f^{(l)WXY} + \a_{kl} \left|f^{(k)}\right|^2 \left|f^{(l)}\right|^2 = 0,
\end{eqnarray}

\vspace{0.64cm}

\begin{table}[h]
\label{table1}
\begin{center}
\begin{tabular}{|l|c|c|c|c|}
\hline
\multicolumn{1}{|c|}{{\bf Group}} & \multicolumn{4}{|c|}{{\bf Embedding}} \\
\hline
$SO(3) \times SO(2,1)^3$ & + & + & + & + \\
\hline
$SO(3) \times SO(2,1)^3$ & - & - & - & - \\
\hline
$SO(3)^2 \times SO(2,1)^2$ & + & - & + & - \\
\hline
$SO(3)^4$ & + & + & - & - \\
\hline
$SO(2,1)^4$ & + & + & - & - \\
\hline
\end{tabular}
\caption{List of $N=4$ semisimple gauge groups involving four simple factors and their embeddings into the fundamental representation of $SO(6,6)$.}
\label{tabla1}
\end{center}
\end{table}

\noindent for $[i,j,k,l]=[1,2,3,4],[1,3,2,4],[1,4,2,3]$. The determinant of Eqs.~(\ref{sys1}) and (\ref{sys2}) identically vanishes, i.e.,
\begin{eqnarray}
\label{determ}
\left|f^{(i)}\right|^2 \left|f^{(j)}\right|^2\left|f^{(k)}\right|^2 \left|f^{(l)}\right|^2 = \frac{1}{3!^4} \left[ \epsilon_{MNPQRSTUVWXY} f^{(i)MNP} f^{(j)QRS} f^{(k)TUV} f^{(l)WXY} \right]^2,
\end{eqnarray}
since, for both groups $SO(3)$ and $SO(2,1)$, $f^{(i)MNP}=f^{(i)}_{MNP}=\epsilon_{MNP}$, thus $\left|f^{(i)}\right|^2= 3!$, when the embedding respects the signature of the metric, and $f^{(i)MNP}=-f^{(i)}_{MNP}=-\epsilon_{MNP}$ and $\left|f^{(i)}\right|^2= -3!$ if the signature is inverted. This proves, in turn, that Eqs.~(\ref{sys1}) and (\ref{sys2}) are consistent with a nontrivial assignment of the de Roo-Wagemans phases.

System (\ref{sys1}) and (\ref{sys2}) explicitly reduces to the following set of  
equations:
\begin{align}
\label{system1}
& \sin(\a_2-\a_1) = \sin(\a_3-\a_4), \\
\label{system2}
& \sin(\a_3-\a_1) = \pm \sin(\a_4-\a_2), \\
\label{system3}
& \sin(\a_4-\a_1) = \pm \sin(\a_2-\a_3),
\end{align}
where the undetermined sign in the right-hand side of Eqs.~(\ref{system2}) and (\ref{system3}) must be set to $+$ when considering the first two gauge groups in Table \ref{tabla1} and it is $-$ for the last three gaugings\footnote{For the third group listed in Table \ref{tabla1} we have permuted the second and third factors for yielding the same system.}. The solutions for the former situation show that all $SL(2)$ vectors are aligned, namely, the group $SO(3) \times SO(2,1)^3$ implies a purely electric gauging. It is only for the latter three gauge groups that at least two de Roo-Wagemans phases can be differently chosen. Actually, under these conditions, there must be the case that two pairs of either parallel or orthogonal $SL(2)$ vectors polarize the decomposition of $G$. In Table \ref{tabla2} we list all allowed duality angles relative to $\alpha_1$.

\vspace{0.57cm}

\begin{table}[h]
\begin{center}
\begin{tabular}{|c|c|c|c|}
\cline{2-4}
\multicolumn{1}{c|}{} & $\a_2$ & $\a_3$ & $\a_4$ \\
\hline
Case A. & $\a$ & $\a$ & $0$ \\
\hline
Case B. & $\pi+\a$ & $0$ & $\a$ \\
\hline
Case C. & $\pi/2$ & $\a+\pi/2$ & $\a$ \\
\hline
Case D. & $-\pi/2$ & $\a$ & $\a+\pi/2$ \\
\hline
\end{tabular}
\caption{Duality angles (relative to $\a_1$) for all nonpurely electric semisimple $N=4$ gaugings compatible with a truncation from $N=8$.}
\label{tabla2}
\end{center}
\end{table}

It can be explicitly checked that the $SL(2)$ angles assignments for the gauge groups in Table~2 do not only solve Eqs.~(\ref{sys1}) and (\ref{sys2}) but the entire system of constraints. In order to understand the reason why this is the case, it will be useful to take a more geometrical point of view for discussing Eq.~(\ref{quad3}). Instead of the real embedding tensor components, let us introduce the complex fluxes $F_{MNP}=f_{+MNP}+if_{-MNP}$, in terms of which Eq.~(\ref{QC46bb2}) can be rewritten as
\beq
\label{rew}
\left[\overline{F} \wedge F \right]_{SD} = 0,
\eeq
where we have used that $\left(\overline{F} \wedge F\right)_{MNPQRS}=\frac{6!}{3!^2} \overline{F}_{[MNP} F_{QRS]}$. Sch\"on-Weidner ansatz is now given by $F=\sum_{k=1}^K e^{i\alpha_k} F^{(k)}$ with obvious definitions, while Eq.~(\ref{rew}) gives
\beq
\label{rew2}
\sum_{k=1}^K  \left[\overline{F}^{(k)} \wedge F^{(k)}\right]_{SD} - 2 i \sum_{k<l=1}^K \mbox{Re} \left[ i e^{i(\a_l-\a_k)} \overline{F}^{(k)} \wedge F^{(l)} \right]_{SD} = 0,
\eeq
where we have used that $\overline{F}^{(k)} \wedge F^{(l)}$ is a $6$-form valued anti-Hermitian matrix for writing the sum of all its entries as the sum of its diagonal elements plus the imaginary parts of its off-diagonal ones.

All assignments listed in Table \ref{tabla2} correspond to a complex flux involving strictly two terms, namely,
\beq
\label{SchWcomplex}
F=e^{i\alpha_1}\left(F^{(1)}+e^{i\alpha} F^{(2)}\right),
\eeq
each one nontrivially defined in a $6$-dimensional space. The expressions for $F^{(1)}$ and $F^{(2)}$ for each solution are listed in Table \ref{tabla3}.

\vspace{0.64cm}

\begin{table}[h]
\begin{center}
\begin{tabular}{|c|c|c|c|c|}
\cline{2-5}
\multicolumn{1}{c|}{} & $F^{(1)}$ & $F^{(2)}$ & ${*F}^{(1)}$ & ${*F}^{(2)}$ \\
\hline
Case A. & $f^{(1)} + f^{(4)}$ & $f^{(2)} + f^{(3)}$ & $-f^{(1)} - f^{(4)}$ & $-f^{(2)} - f^{(3)}$ \\
\hline
Case B. & $f^{(1)} + f^{(3)}$ & $- f^{(2)} + f^{(4)}$ & $-f^{(1)} - f^{(3)}$ & $f^{(2)} - f^{(4)}$ \\
\hline
Case C. & $f^{(1)} + if^{(2)}$ & $if^{(3)} + f^{(4)}$ & $if^{(1)} - f^{(2)}$ & $ - f^{(3)} + if^{(4)}$ \\
\hline
Case D. & $f^{(1)} - i f^{(2)}$ & $f^{(3)} + if^{(4)}$ & $if^{(1)} + f^{(2)}$ & $if^{(3)} - f^{(4)}$ \\
\hline
\end{tabular}
\caption{Decompositions of all semisimple gaugings solving (\ref{quad3}) in terms of two $3$-forms effectively defined on orthogonal $6$-dimensional.}
\label{tabla3}
\end{center}
\end{table}

When a decomposition like Eq.~(\ref{SchWcomplex}) holds, Eq.~(\ref{rew2}) reduces to
\beq
\label{rew3}
\left[\overline{F}^{(1)} \wedge F^{(1)}\right]_{SD} + \left[\overline{F}^{(2)} \wedge F^{(2)}\right]_{SD} - 2 i \mbox{Re} \left[ i e^{i\a} \overline{F}^{(1)} \wedge F^{(2)} \right]_{SD} = 0.
\eeq
This identity should be satisfied for every choice of $\a$ so that both of its terms must separately vanish. Thus we get the following set of phase-independent equations:
\beq
\label{rew4}
\left[\overline{F}^{(1)} \wedge F^{(1)}\right]_{SD} + \left[\overline{F}^{(2)} \wedge F^{(2)}\right]_{SD} = 0,
\eeq
\beq
\label{rew5}
\left[\overline{F}^{(1)} \wedge F^{(2)} \right]_{SD} = 0.
\eeq

Let $*F^{(1)}$ and $*F^{(2)}$ be the trivial extensions of the Hodge-duals of $F^{(1)}$ and $F^{(2)}$ relative to their $6$-dimensional domains\footnote{Notice that the referred star-operators are defined on different spaces.}, respectively. Using the fact that the $12$-dimensional Levi-Civita symbol effectively acts on the direct product of these subspaces as the product of the corresponding $6$-dimensional Levi-Civita symbols, it is straightforward to prove that
\beq
\label{unamas}
{*\left( \overline{F}^{(1)} \wedge F^{(2)} \right)} = - \left({*\overline{F}}^{(1)} \wedge {*F}^{(2)}\right),
\eeq
therefore, Eq.~(\ref{rew5}) can be equivalently written as
\beq
\label{quad3bisb}
\overline{F}^{(1)} \wedge F^{(2)} = {*\overline{F}}^{(1)} \wedge {*F}^{(2)}.
\eeq
For the first two cases in Table \ref{tabla3} we have $*F^{(1)}=-F^{(1)}$ and $*F^{(2)}=-F^{(2)}$, while for the third and the fourth cases we have  $*F^{(1)}=-iF^{(1)}$ and $*F^{(2)}=-iF^{(2)}$, so that Eq.~(\ref{quad3bisb}) is clearly satisfied. We stress that we have been able to write Eq.~(\ref{unamas}) because of the decomposition of $F$ in two terms with $6$-dimensional orthogonal domains. For the former situation Eq.~(\ref{rew4}) becomes trivial since both $F^{(1)}$ and $F^{(2)}$ are real forms and the wedge product is antisymmetric. For the latter cases we find $*\left( \overline{F}^{(1)} \wedge F^{(1)} \right) = - \overline{F}^{(2)} \wedge F^{(2)}$ and therefore $*\left( \overline{F}^{(2)} \wedge F^{(2)} \right) = - \overline{F}^{(1)} \wedge F^{(1)}$. It follows that Eq.~(\ref{rew4}) also holds for these cases.

Concerning constraint (\ref{QC46bbbo}), the situation resembles the already discussed case of purely electric gaugings. Indeed, for the first and second duality phase assignments in Table \ref{tabla3} the preimages of $\eta_{MN}$ under the embeddings associated with the same de Roo-Wagemans angle differ in a sign and, therefore, the contributions to $f_{\a MNP} f_{\b}^{MNP}$ coming from both factors cancel, i.e., those contributions to the left-hand side of Eq.~(\ref{QC46bbbo}) coming from factors with different $SL(2)$ angles separately vanish. For the last two cases in Table \ref{tabla3} the situation is quite similar, the only difference being that the contribution to Eq.~(\ref{QC46bbbo}) coming from factors that do not share the same duality angle cancel mutually. In any case, every solution of Eq.~(\ref{quad3}) proves to be also a solution of Eq.~(\ref{QC46bbbo}), making this last constraint redundant with respect to Eq.~(\ref{quad3}) within the context of the de Roo-Wagemans formalism.

Before discussing the moduli stability for these semisimple gauge groups, let us notice that many nonsemisimple realizations can also be constructed for consistently gauging a halved maximal supergravity using the Sch\"on-Weidner ansatz. As a particular example let us mention $\left(U(1)^3\right)^2\times SO(3)^2$ with the $3$-form associated with $U(1)^3$ being the one introduced when analyzing purely electric gaugings.

\subsection{Stability analysis}

The study of the scalar potential and the mass matrix associated with the fluxes we have found in the previous section is greatly simplified due to the exhaustive study of semisimple gaugings for $N=4$ supergravity performed in Refs.~\cite{deRoo:2002jf, deRoo:2003rm}. Let us briefly quote the main results.

Once the potential is extremized in the $SL(2)$ scalar sector, it takes the following form:
\beq
\label{potential}
V(\nu)=\frac{C}{\left|C\right|}\,\sqrt{\Delta} - T,
\eeq
where
\beq
\label{C}
C=\sum_{i,j=1}^K \cos(\a_j-\a_i) V_{ij},
\eeq
\beq
\label{Delta}
\Delta=2\sum_{i,j,k,l=1}^K \a_{ik}\a_{jl}V_{ij}V_{kl},
\eeq
\beq
\label{T}
T=-\sum_{i,j=1}^K \a_{ij} W_{ij},
\eeq
with
\beq
\label{V}
V_{ij}=\frac{1}{4}\left[\nu^{MQ}\nu^{NR}\left(\eta^{PS}+\nu^{PS}\right)-\frac{1}{3} \nu^{MQ}\nu^{NR}\nu^{PS}\right] f^{(i)}_{MNP} f^{(j)}_{QRS},
\eeq
\beq
\label{W}
W_{ij}=\frac{1}{36} \,\epsilon^{abcdef} \nu_a{}^{M}\nu_b{}^{N}\nu_c{}^{P}\nu_d{}^{Q}\nu_e{}^{R}\nu_f{}^{S} f^{(i)}_{MNP} f^{(j)}_{QRS}.
\eeq
In these equations, $\nu_a{}^{M}$ corresponds to the six matter multiplets in the theory and, by definition, $\nu^{MQ}=\nu_a{}^{M}\nu_a{}^{Q}$. The condition for an extremum to exist is that $\Delta>0$ which precisely implies that at least two duality angles must be different.

The potential is analyzed at the origin of the scalar manifold, namely, for $\nu_{0a}{}^M=1$ when $M$ refers to a noncompact direction and $\nu_{0a}{}^M=0$ otherwise. Under these conditions, one has $V_{0ij}=0$ for $i\ne j$ and the previous expressions simplify to
\beq
\label{C0}
C_0=\sum_{i=1}^K V_{0ii},
\eeq
\beq
\label{Delta0}
\Delta_0=2\sum_{i,j=1}^K \a_{ij}^2 V_{0ii}V_{0jj},
\eeq
\beq
\label{T0}
T_0=-\sum_{i,j=1}^K \a_{ij} W_{0ij},
\eeq
while the explicit values of $V_{0ii}$ and $W_{0ij}$ are:
\beq
\label{V0}
V_{0ii}=\left\{ 
\begin{array}{ll}
	-\frac{1}{2} & \mbox{for} \,\, SO(3)_-, \\
	\frac{1}{2} & \mbox{for} \,\, SO(2,1)_+, \\
	0 & \mbox{for} \,\, SO(3)_+ \,\, \mbox{and} \,\, SO(2,1)_-,
\end{array}
\right.
\eeq
\beq
\label{W0}
W_{0ij}=\left\{ 
\begin{array}{ll}
	1 & \mbox{for} \,\, SO(3)_-, \\
	0 & \mbox{for} \,\, SO(3)_+,\,SO(2,1)_+ \,\, \mbox{and} \,\, SO(2,1)_-.
\end{array}
\right.
\eeq
The subindices refer to the respective embeddings.

For the third gauge group listed in Table \ref{tabla1} we find $\Delta_0<0$ so that the potential does not exhibit any extremum in this case. We are left with the gauge groups $SO(3)^4$ and $SO(2,1)^4$.

\subsubsection{$SO(3)^4$}

In this case we obtain $C_0=-1$, $\Delta_0=a^2$, $T_0=2a$ and $V_0=-|a|-2a$, where $a=\pm\sin\a, \pm 1$ for the duality angles assignments in Table \ref{tabla2}, respectively. The eigenvalues of the mass matrix\footnote{We refer to Ref.~\cite{deRoo:2003rm} for the expressions of the first and second derivatives of the potential and the mass matrix.} for the $36$ matter scalars all equal $-2a$, and then, for a de Sitter solution, these eigenvalues are all positive. Since $C_0<0$, it follows that for the $SL(2)$ scalars the potential exhibit a maximum, namely, there are two tachyons present in the $SL(2)$ sector.

\subsubsection{$SO(2,1)^4$}

Now we have $C_0=1$, $\Delta_0=a^2$, $T_0=0$ and $V_0=|a|$, where $a=\pm\sin\a, \pm 1$ for the cases listed in Table \ref{tabla2}, respectively. Eight eigenvalues of the mass matrix identically vanish. They correspond to the Goldstone bosons producing the masses of the gauge fields after the breaking of the local symmetry to $U(1)^4$. There are always two negative eigenvalues equal to $-2|a|$, and setting $a=0$ does not help since this case reduces to a purely electric gauging.  

\section{Concluding remarks}

When halving maximal $D=4$ gauged supergravity, in addition to the usual constraints to the gauge groups embeddings of half-maximal models, other ones arise as a consequence of the truncation from $N=8$ to $N=4$. In this paper we have explored the possibility of solving these new constraints by using the only known semianalytical approach to the problem, namely, the de Roo-Wagemans phases method. We have performed the implementation of this procedure within the context of the embedding tensor formalism by means of the Sch\"on-Weidner ansatz.

We have proven that no semisimple gaugings, with the eventual exception of those associated with groups that decompose in exactly four simple $3$-dimensional factors, allow the stabilization of all moduli since the truncation constraints force the duality angles to be equal or differ by multiples of $\pi$, i.e., semisimple choices with less than four simple factors reduce to purely electrical gaugings.

For the five semisimple groups constructed upon four copies of $SO(3)$ and/or $SO(2,1)$, the only $3$-dimensional simple subgroups of $SO(6,6)$, we have determined the $SL(2)$ phases that are compatible with all the constraints. Only three of them can have two or more different duality angles, showing that the impact of halving maximal supergravity within the context of the de Roo-Wagemans deformation is a drastic reduction of up to $85\%$ of those $N=4$ semisimple gaugings susceptibles of a nontrivial stability analysis and a highly restricted assignment of the duality phases: only two pairs of either parallel or orthogonal $SL(2)$ vectors can deform the decompositions of the corresponding gauge groups. Interestingly, we have realized that all the solutions of one of the truncation constraints automatically become solutions of the other, making this last one redundant in the context of the de Roo-Wagemans method. We have also shown that these solutions admit a suitable geometrical interpretation since they can be characterized in terms of two $3$-forms nontrivially defined on mutually orthogonal $6$-dimensional spaces that are separately self-dual or anti-self-dual relative to their domains. 

As a final point, we have discussed the scalar potential and mass matrix properties near the origin of the scalar manifold for the resulting models. While one of them does not present any extremum, the other two do exhibit extrema, although not fully stable under fluctuations of all scalar fields. It would be interesting to further analyze nonsemisimple gaugings using similar techniques as the ones we have employed in order to investigate up to what extent it is possible to enhance the chances for moduli stabilization when considering halved maximal supergravity models.

\section*{Acknowledgments}

We are grateful to C.~N\'u\~nez for her valuable insights during all stages of this project. We thank D.~Marqu\'es for carefully reading the manuscript. We also thank G.~Aldaz\'abal and M.~Gra\~na for useful discussions. This work has been partially supported by projects No.~UBACyT 20020100100669, No.~PIP 11220080100507 and No.~PICT-2007-02182.

\end{document}